\documentclass[preprint,aps,prd,showpacs,nofootinbib]{revtex4}

\usepackage{graphicx}
\usepackage{amsfonts}

\newcommand*{\chpt}{\raise0.4ex\hbox{$\chi$}PT}
\newcommand*{\schpt}{S\raise0.4ex\hbox{$\chi$}PT}
\newcommand*{\ie}{\textit{i.e.},\ }
\newcommand*{\eg}{\textit{e.g.},\ }
\newcommand*{\etc}{\textit{etc.}}

\newcommand*{\et}{\textit{et al.}}

\renewcommand*{\prd}[1]{Phys.\ Rev.\ \textbf{D#1}}
\newcommand*{\plb}[1]{Phys.\ Lett.\ \textbf{#1 B}}
\newcommand*{\npbps}[1]{Nucl.\ Phys.\ \textbf{B}
	(Proc.\ Suppl.) \textbf{#1}}

\newcommand*{\boston}{Nucl.\ Phys.\ {\bf B}
 (Proc.\ Suppl.) {\bf 119} (2003)}

\newcommand*{\MeV}{{\rm Me\!V}}

\newcommand*{\Tr}{\textrm{Tr}}

\newcommand{\cD}{\mathcal{D}}

\newcommand{\cI}{\mathcal{I}}

\newcommand{\cL}{\mathcal{L}}
\newcommand{\cM}{\mathcal{M}}

\newcommand{\cO}{\mathcal{O}}
\newcommand{\cP}{\mathcal{P}}

\newcommand{\cU}{\mathcal{U}}
\newcommand{\cV}{\mathcal{V}}

\def\eq#1{Eq.~(\ref{eq:#1})}
\def\eqs#1#2{Eqs.~(\ref{eq:#1}) and (\ref{eq:#2})}
\def\eqsthru#1#2{Eqs.~(\ref{eq:#1}) through (\ref{eq:#2})}
\def\eqsor#1#2{Eq.~(\ref{eq:#1}) or (\ref{eq:#2})}

\begin{document}

\title{Pseudoscalar Decay Constants in Staggered
Chiral Perturbation Theory}
\author{C. Aubin}
\author{C. Bernard}
\affiliation{Washington University, St. Louis, MO 63130}
\begin{abstract}
In a continuation of an ongoing program,
we use staggered chiral perturbation theory to 
calculate the one-loop chiral logarithms and analytic terms 
in the pseudoscalar meson leptonic 
decay constants, $f_{\pi^+_5}$ and $f_{K^+_5}$.
We consider the partially quenched, ``full QCD'' (with three
dynamical flavors), and quenched cases.
\end{abstract}
\pacs{12.39.Fe, 11.30.Rd, 12.38.Gc}

\maketitle

\section{Introduction}\label{sec:intro}

Simulations with staggered (Kogut-Susskind, KS) fermions are  very 
fast relative to other available approaches, making possible
simulations of QCD that include the effects of
light sea quarks \cite{CHIRAL_PANEL_2001}. 
However, with currently practical lattice spacings (\eg MILC simulations 
\cite{PETER_PRL,IMP_SCALING,IMP_SCALING2,MILC_SPECTRUM,FB_LAT02}
at $a\approx 0.09 - 0.13\ \rm fm$) 
taste\footnote{We use \cite{AUBIN,LAT02} the
term ``taste'' to denote the different KS species resulting from doubling,
and ``flavor'' for the physical 
$u$-$d$-$s$ quantum number.} violations are not negligible. 
Thus fits to such lattice data should take into account
the taste-violating effects; indeed, if such effects are not
taken into account, the speed advantage of KS fermions may be
offset by the size of the systematic errors. The
taste-violating effects can be calculated in 
a systematic way using staggered chiral
perturbation theory (\schpt).  

In Ref.~\cite{AUBIN}, we formulate
\schpt\ for the physical case of multiple flavors.  \schpt\ is then used to 
calculate the one-loop chiral logarithms in the pion and kaon masses.
Here, we continue the program of Ref.~\cite{AUBIN} and compute 
$f_{\pi^+_5}$ and
$f_{K^+_5}$, the $\pi^+$ and $K^+$ leptonic decay constants for the Goldstone
mesons, to one loop. 
As we have laid most of the
necessary groundwork already, we will merely state what is necessary
for this present work and refer the reader to Ref.~\cite{AUBIN} for
the details common to both calculations. 
As in the calculation of the $\pi^+$ and $K^+$ masses, we perform our
calculation using three dynamical KS-flavors (each with four tastes),
which we call the $4\!+\!4\!+\!4$ theory, and later adjust the result by hand
using a quark flow technique \cite{AUBIN,SHARPE_QCHPT} to a
$1\!+\!1\!+\!1$ theory (three flavors each with a single taste).

The outline of this paper is as follows: In Sec.~\ref{sec:ls-lag}, we
write down the \schpt\ Lagrangian for three dynamical flavors.  We then
calculate, in Sec.~\ref{sec:fpi_fK}, the one-loop chiral logarithms
which contribute to the flavor-nonsinglet Goldstone meson decay
constant in the partially quenched case. 
Here we keep three dynamical
flavors but add two additional quenched flavors as valence quarks,
which in the general case have distinct masses from the
dynamical (sea) quarks. The transition to a $1\!+\!1\!+\!1$ theory is then 
made. There are only a few differences in this
procedure from that of Ref.~\cite{AUBIN}. We also give the results in
the fully quenched case. 
The full next-to-leading order (NLO) results, including  the analytic terms,
are presented in Sec.~\ref{sec:final_results} for various relevant cases.
We conclude with some comments in Sec.~\ref{sec:conc}. An Appendix
gives technical details about the evaluation of the one-loop
integrals that arise in Sec.~\ref{sec:fpi_fK}.

\section{The Lee-Sharpe Lagrangian for 3 flavors}
\label{sec:ls-lag}

The starting point for \schpt\ is the Lee-Sharpe Lagrangian
\cite{LEE_SHARPE} generalized to multiple flavors.
In Ref.~\cite{AUBIN} we examined a general $n$-flavor
theory\footnote{Here $n$ refers to 
the number of sea quarks.} and later
specialized to $n=3$. Here we take $n=3$ from
the beginning. For 3 KS flavors, $\Sigma=\exp(i\Phi / f)$ is a $12
\times 12$ matrix, with $\Phi$ given by:
\begin{eqnarray}\label{eq:Phi}
	\Phi = \left( \begin{array}{ccc}
     		 U  & \pi^+ & K^+  \\*
     		 \pi^- & D & K^0   \\*
		 K^-  & \bar{K^0}  & S\end{array} \right),
\end{eqnarray}
where $U = \sum_{a=1}^{16} U_a T_a$ 
(and similarly for $\pi^+$, $K^+$, \dots),  
with
\begin{equation}\label{eq:T_a}
	T_a = \{ \xi_5, i\xi_{\mu5}, i\xi_{\mu\nu}, \xi_{\mu}, \xi_I\}.
\end{equation}
We use the Euclidean gamma matrices $\xi_{\mu}$, with
$\xi_{\mu\nu}\equiv \xi_{\mu}\xi_{\nu}$ ($\mu <\nu$ in \eq{T_a}),
$\xi_{\mu5}\equiv \xi_{\mu}\xi_5$, and $\xi_I \equiv I$ is the
$4\times 4$ identity matrix. The field $\Sigma$ transforms under
$SU(12)_L\times SU(12)_R$ as $\Sigma \rightarrow L\Sigma R^{\dagger}$.
The component fields of the diagonal (flavor-neutral) elements ($U_a$,
$D_a$, and $S_a$) are real; the other, charged, fields 
($\pi^+_a$, $K^0_a$, \etc) are complex,
so that $\Phi$ is Hermitian. The mass
matrix is given by the $12\times 12$ matrix
\begin{eqnarray}
	\cM = \left( \begin{array}{ccc}
     		m_u I  & 0 &0  \\*
     		0  & m_d I & 0   \\*
		0  & 0  & m_s I   \end{array} \right)\ .
\end{eqnarray}

Our (Euclidean) Lagrangian is:
\begin{eqnarray}\label{eq:final_L}
	\cL & = & \frac{f^2}{8} \Tr(\partial_{\mu}\Sigma 
	\partial_{\mu}\Sigma^{\dagger}) - 
	\frac{1}{4}\mu f^2 \Tr(\cM\Sigma+\cM\Sigma^{\dagger})
	+ \frac{2m_0^2}{3}(U_I + D_I + S_I)^2 + a^2 \cV,
\end{eqnarray}
where $\mu$ is a constant with units of mass, 
$\Tr$ is the full $12\times 12$ trace, and $\cV=\cU+\cU\,'$ is
the taste-symmetry breaking potential. The $\cU$ term is given in 
Ref.~\cite{AUBIN}; it is not needed explicitly here.
For $\cU\,'$, we have
\begin{eqnarray}
	\label{eq:U_prime}
	-\cU\,'  & = & C_{2V}\frac{1}{4} 
		\sum_{\nu}[ \Tr(\xi^{(3)}_{\nu}\Sigma)
	\Tr(\xi^{(3)}_{\nu}\Sigma)  + h.c.] \nonumber \\*
	&&+C_{2A}\frac{1}{4} \sum_{\nu}[ \Tr(\xi^{(3)}_{\nu
         5}\Sigma)\Tr(\xi^{(3)}_{5\nu}\Sigma)  + h.c.] \nonumber \\*
	& & +C_{5V}\frac{1}{2} \sum_{\nu}[ \Tr(\xi^{(3)}_{\nu}\Sigma)
	\Tr(\xi^{(3)}_{\nu}\Sigma^{\dagger})]\nonumber \\*
	& & +C_{5A}\frac{1}{2} \sum_{\nu}[ \Tr(\xi^{(3)}_{\nu5}\Sigma)
	\Tr(\xi^{(3)}_{5\nu}\Sigma^{\dagger}) ],
\end{eqnarray}
where the $\xi^{(3)}_B$
are block-diagonal $12\times 12$ matrices:
\begin{equation}\label{eq:xi_B}
	\xi^{(3)}_B = \left(\matrix{\xi_B & 0 & 0 \cr
	0 & \xi_B & 0 \cr
	0 & 0 & \xi_B \cr}\right)\ ,
\end{equation}
with $\xi_B$ the $4\times 4$ objects,
and $B\in\{5,\mu,\mu\nu\ (\mu<\nu),\mu5,I \}$.

\begin{figure}
 \includegraphics[width=3in]{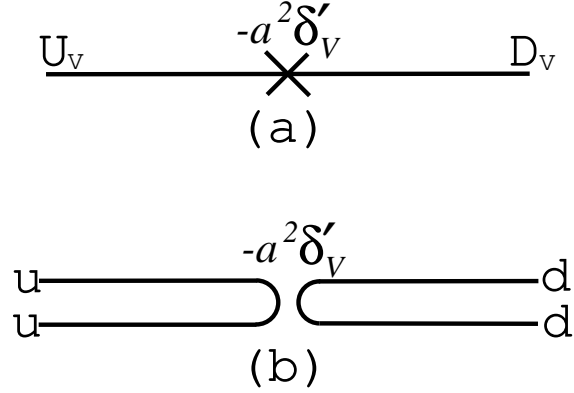} \caption{The two-point
  	mixing vertex coming from the $\cU\,'$ term. (a)
  	corresponds to the chiral theory (we also have similar $U-S$
  	and $D-S$ mixing terms). (b) shows the corresponding quark level
        diagram.  Here we only show the mixing among
  	the taste-vectors, but there are similar vertices among the
  	axial tastes, as well as the singlet tastes (with $a^2\delta'_V \to
        4m_0^2/3)$.}  \label{fig:2-pt_vertex}
\end{figure}

As seen in Ref.~\cite{AUBIN}, $\cU\,'$ generates two-point vertices 
at $\cO(a^2)$ (shown in
Fig.~\ref{fig:2-pt_vertex}) that mix flavor-neutral particles of 
vector and axial tastes. In addition, flavor-neutral, taste-singlet
particles are mixed by the $m_0^2$ term in \eq{final_L}, which
results from the anomaly. In all three cases (taste vector, axial vector,
and singlet), we have a term in
the Lagrangian of the form $(\delta'/2) (U + D + S)^2$, where
\begin{equation}\label{eq:dp_def}
	\delta' = \cases { a^2 \delta'_V, &taste-vector;\cr
	a^2 \delta'_A, &taste-axial;\cr
	4m_0^2/3, &taste-singlet.\cr }
\end{equation}
Expressions for $\delta'_V$ and $\delta'_A$
in terms of the coefficients of 
$\cU\,'$ are given in Ref.~\cite{AUBIN}. 
These mixings require us to diagonalize the full mass matrix in each of the
three channels. 
We write the
propagator for the vectors as:
\begin{eqnarray}\label{eq:prop}
	G_V& =& G_{0,V} + \cD^V \ .
\end{eqnarray}
$\cD^V$ is the part of the taste-vector flavor-neutral propagator that
is disconnected at the quark level (\ie Fig.~\ref{fig:2-pt_vertex}
plus iterations of intermediate sea quark loops).  Explicitly,
we have \cite{AUBIN}:
\begin{eqnarray} \label{eq:D_term}
	\cD^V_{MN}&=&  -a^2\delta'_V 
	\frac{ (q^2 + m_{U_V}^2)(q^2 + m_{D_V}^2)(q^2 + m_{S_V}^2) }
	{(q^2 + m_{M_V}^2)(q^2 + m_{N_V}^2)
	(q^2 + m_{\pi^0_V}^2)(q^2+m_{\eta_V}^2)(q^2 + m_{\eta'_V}^2)}\ .
\end{eqnarray}
Here, $m_{\pi^0_V}^2$, $m_{\eta_V}^2$ and $m_{\eta'_V}^2$ are the
eigenvalues of the full mass-squared matrix (\ie the poles of
$G_V$).  
We emphasize that \eq{D_term} remains valid in the $n=3$ partially quenched 
case. 
The external mesons $M$ and $N$ may be any
flavor-neutral states, made from either sea quarks or valence quarks.

In the quenched case $\cD^V_{MN}$ is simply
\begin{equation}\label{eq:D_term_quenched}
	\cD^{V,\rm quench}_{MN}= -a^2\delta'_V 
		\frac{1}{(q^2 + m_{M_V}^2)(q^2 + m_{N_V}^2)}\ .
\end{equation}

Equations~(\ref{eq:prop}) through (\ref{eq:D_term_quenched}) apply
explicitly to the taste-vector channel; to get the result in
the taste-axial (taste-singlet) 
channel, just let $V\to A$ ($V\to I$ and $a^2\delta'_V\rightarrow 4m^2_0/3$).  
In the quenched case we cannot take $m_0\to\infty$, and must include
additional $\eta'_I$-dependent terms in the 
Lagrangian, resulting in the replacement $m_0^2\to
m_0^2 + \alpha q^2$ in the singlet form of 
\eq{D_term_quenched} \cite{AUBIN,CBMG_QCHPT}. 

\section{One loop decay constant for 4+4+4 dynamical flavors}
\label{sec:fpi_fK}

We calculate the pion\footnote{We refer generically to
any flavor-charged meson as a ``pion.''} decay constant
in a partially quenched theory. Full theory results are easily obtained
by taking appropriate limits.
There are three sea quarks ($u$, $d$
and $s$) and two valence quarks ($x$ and $y$). The pion of interest
is the $P^+_5$, a Goldstone pion which is composed of an $x\bar{y}$
pair of quarks. 

The $P^+_5$ decay constant is defined by the matrix
element:
\begin{equation}\label{eq:matrix_element}
	\left\langle 0 \left| j_{\mu5}^{P^+_5}
	\right| P^+_5(p) \right\rangle =-i
	f_{P^+_5} p_{\mu} \ ,
\end{equation}
where $j_{\mu5}^{P^+_5}$ is the axial current corresponding to
${P^+_5}$.
With this normalization, $f_{\pi} \approx 131\ \MeV$.
In terms of $\Sigma$, we can write the
axial current as
\begin{eqnarray}\label{eq:current}
	j_{\mu5}^{P^+_5} &= &\frac{-i f^2}{8} 
	   	\Tr\left[\xi^{(3)}_5 \cP^{P^+} \left(
	   	\partial_{\mu}\Sigma\Sigma^{\dagger} +
	   	\Sigma^{\dagger}\partial_{\mu}\Sigma
		\right)\right]\ .
\end{eqnarray}
Here $\cP^{P^+}$ projects out the $4\times 4$ block
with appropriate flavor: 
If we make $x$ and $y$ the last
two flavors of $\Sigma$, then  $\cP^{P^+}_{ij} = \delta_{i5}\delta_{j4}$,
where $i,j$ are flavor indices.

\begin{figure}
 \includegraphics[width=3in]{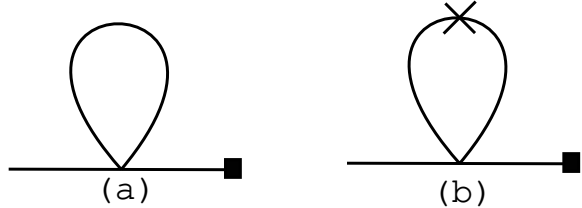} \caption{The \schpt\ diagrams
	contributing to the pion decay constant, coming from
	wave-function renormalization. 
	The box represents the the axial current. 
	(a) is the connected piece,
	where the propagator in the loop contains no two-point vertex
	insertions. (b) subsumes the graphs which have disconnected
	insertions within the loop. The cross represents one or more
	insertions of the $\delta'$ vertex, with $\delta'$ given in
	\eq{dp_def}.}  \label{fig:tadZ}
\end{figure}

\begin{figure}
 \includegraphics[width=3in]{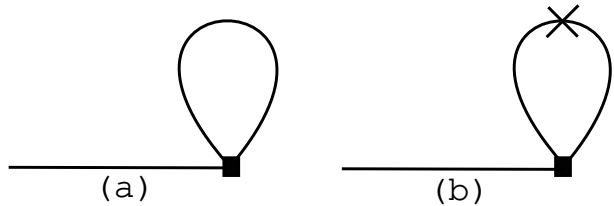} \caption{Same as 
	Fig.~\ref{fig:tadZ}, but these
	contributions to the decay constant are from axial current
	corrections.}  \label{fig:tad_f}
\end{figure}

At one loop, the decay constant has the form 
\begin{equation}\label{eq:fpi}
	f_{P^+_5} = f\left( 1 + \frac{1}{16\pi^2 f^2} \delta\!
	f_{P^+_5}\right) \ .
\end{equation}
There are two contributions to $\delta\! f_{P^+_5}$,
which we call $\delta\! f^{\rm Z}_{P^+_5}$
and $\delta\! f^{\rm current}_{P^+_5}$.  They are shown in
Figs.~\ref{fig:tadZ} and \ref{fig:tad_f}, respectively.
The contribution $\delta\! f^{\rm Z}_{P^+_5}$ is merely 
wavefunction renormalization. We have:
\begin{equation}\label{eq:wave_ren_contr}
	\delta\! f^{\rm Z}_{P^+_5} = \frac{1}{2}\delta\! Z_{P^+_5}
	\equiv -\frac{16\pi^2f^2}{2}\;
	       {d\Sigma(p^2) \over dp^2}  \ .
\end{equation}
The self-energy, $\Sigma(p^2)$, has already been
calculated in Ref.~\cite{AUBIN}. The wavefunction renormalization
arises only from the vertex generated by the kinetic energy term in
\eq{final_L}, since derivatives on the external lines are necessary
to generate $p^2$ dependence in a tadpole diagram.  The factor of $1/2$ in
\eq{wave_ren_contr} is due to the
fact that this diagram is multiplied by $\sqrt{Z} = \sqrt{1+\delta\! Z}
\approx 1 +\frac{1}{2}\delta\! Z$.

The contribution $\delta\! f^{\rm current}_{P^+_5}$ is the current correction.
It arises from the expansion of 
\eq{current} to cubic order in $\Phi$. 
Performing this expansion, it is easy to see that the wavefunction
and current correction terms are proportional to each
other:  $\delta\! f^{\rm current}_{P^+_5} = -4\delta\! f^{\rm
Z}_{P^+_5}$. This fact, noted also in  \cite{SHARPE_QCHPT},
is perhaps not surprising, since the form of the axial current,
\eq{current}, is determined through Noether's theorem only 
by the kinetic energy part of the Lagrangian.
From \eq{wave_ren_contr}, we then have
\begin{eqnarray}\label{eq:deltafpi}
	\delta\! f_{P^+_5}& = &\delta\! f^{\rm Z}_{P^+_5}
	 + \delta\! f^{\rm current}_{P^+_5} = 
	 -\frac{3}{2}\delta\! Z_{P^+_5} \ .
\end{eqnarray}

Using intermediate expressions for $\Sigma(p^2)$ from \cite{AUBIN},
the one-loop result is
\begin{eqnarray}\label{eq:df_result}
	\delta\! f_{P^+_5} & = &
	 -\frac{1}{8}\int\frac{d^4 q}{\pi^2} \Biggl[    
	\sum_{Q,B} \biggl( \frac{1}{q^2 + m^2_{Q_B}}\biggr)
	+ \cD^I_{XX} - 2\cD^I_{XY} +
	 \cD^I_{YY} \nonumber \\*
	&&+  4\cD^V_{XX} +8\cD^V_{XY}
	 + 4\cD^{V}_{YY}+  4\cD^A_{XX} +8\cD^A_{XY}
	 + 4\cD^{A}_{YY}  \Biggr]\ .
\end{eqnarray}
Here, $Q$ runs over the six mesons formed from one
valence and one sea quark (\ie the $xu$, $xd$, $xs$, $yu$, $yd$, and $ys$
mesons). As before, $B$ takes on the 16 values
$\{ 5, \mu, \mu\nu (\mu <\nu), \mu 5, I\}$. 
We have already included the factor of 4 that comes from
summing over the degenerate vector and axial contributions in the
$\cD^V$ and $\cD^A$ terms.
Despite the fact that the only 4-point vertices contributing to this expression
come from the kinetic energy term,
the result is more complicated than that
for the mass renormalization \cite{AUBIN} because 
there are no cancellations here (either accidental or
required by symmetry).

\begin{figure}
 \includegraphics[width=2in]{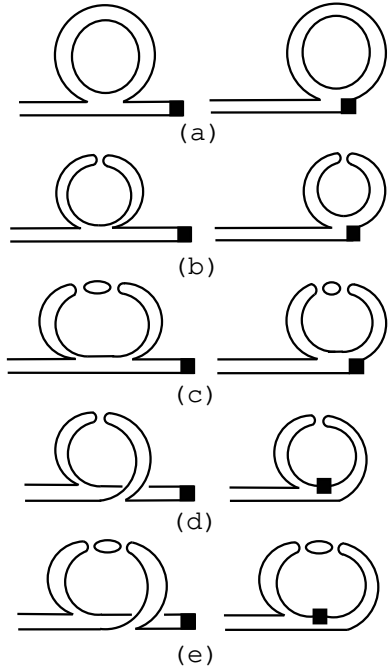} \caption{
        The quark level
  	diagrams that contribute to the one-loop pion decay
  	constant. 
	The box represents an insertion of the axial current. 
	The diagrams on the left correspond to the
  	wavefunction renormalization while those on the right
  	correspond to the current corrections.}  \label{fig:flow}
\end{figure}

\begin{figure}
 \includegraphics[width=3in]{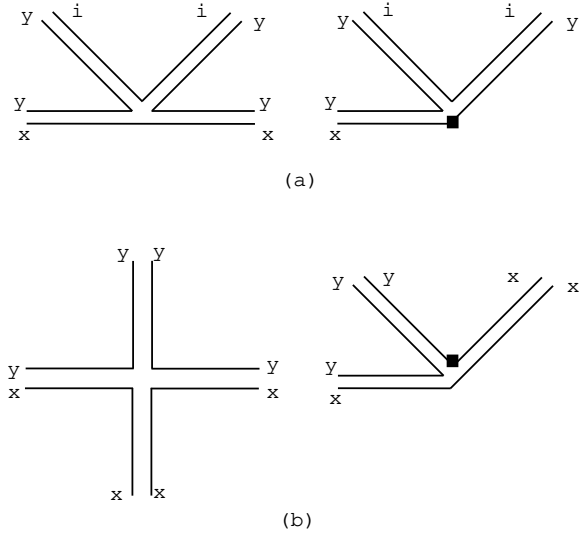} \caption{
	The quark level diagrams for $2\rightarrow 2$ meson scattering which
  	contribute to $f_{P^+_5}$. The indices $i$ and $j$ represent
  	arbitrary quark flavors. There are two additional diagrams
  	(not shown), which are like those in (a) but have the roles of
  	$x$ and $y$ interchanged. The box stands for the axial current.}
  	\label{fig:Vertices}
\end{figure}

The first term \eq{df_result} (the sum over $Q$ and $B$) comes
from the wave function renormalization and current correction diagrams 
shown in Fig.~\ref{fig:flow}(a), which involve a single virtual
quark loop.  The diagrams arise from the vertices in
 Fig.~\ref{fig:Vertices}(a), respectively, where 
$i$ is summed over the sea quarks only. In this case, the propagator in
the loop must be connected since the loop meson is not flavor-neutral.

The vertices in  Fig.~\ref{fig:Vertices}(a) also produce diagrams with
disconnected loop propagators,  Fig.~\ref{fig:flow}(b) and (c), when
$i=y$ (or $i=x$ in the $y\leftrightarrow x$ version of 
Fig.~\ref{fig:Vertices}(a)).  These diagrams give rise to
the $\cD_{YY}$ and $\cD_{XX}$ terms in \eq{df_result}.  

Finally, the
vertices in Fig.~\ref{fig:Vertices}(b) generate the diagrams
Fig.~\ref{fig:flow}(d) and (e). The $\cD_{XY}$  terms in \eq{df_result}
come from these diagrams.
For more discussion of how to identify quark flow diagrams with
the \schpt\ contributions, see Ref.~\cite{AUBIN}.

We can write down the quenched result by (1) eliminating
the term summed over $Q$ and $B$, which arises from virtual
quark loops (diagrams Fig.~\ref{fig:flow}(a)), and 
(2) replacing of $\cD \to \cD^{\rm quench}$ for 
$\cD^V$, $\cD^A$, and $\cD^I$.
These replacements eliminate diagrams
Fig.~\ref{fig:flow}(c) and (e).

In the partially quenched case, 
when the $x$ {\it or} $y$ quark mass is different from
all sea quark masses,
there are double poles in \eq{df_result}
coming from the $\cD_{XX}$ or $\cD_{YY}$ terms.  This is different
from the mass renormalization result, where
double poles do not arise unless $m_x=m_y$ {\it and} this mass is
different from all sea quark masses, a case we did 
not treat in detail in Ref.~\cite{AUBIN}. 
In order to write down explicit results
for partial quenching here,
we must therefore 
expand on the notation of Ref.~\cite{AUBIN}. 
Below we will use the notation defined in the Appendix, 
where we explain the techniques we use to evaluate the integrals.

Before performing the momentum integrals, we
now make the transition from the $4\!+\!4\!+\!4$ case to the
$1\!+\!1\!+\!1$ case. This is easily accomplished since we 
have already determined
the separate contributions from diagrams in Fig.~\ref{fig:flow}
with various numbers of sea quark loops.
Those diagrams with a connected propagator in the
loop, Fig.~\ref{fig:flow}(a), have a single sea quark loop and simply
must be divided by 4.

The remaining diagrams have the same form as those treated in
Ref.~\cite{AUBIN}, so we just briefly review the procedure. 
Diagrams  (b) and (d) in Fig.~\ref{fig:flow} 
have no sea quark loops and are unaffected by the transition
to the $1\!+\!1\!+\!1$ case.  These diagrams have a single factor of
$\delta'$ coming from the overall coefficient
of the disconnected propagator $\cD$ in \eq{D_term}.
Each sea quark loop added on to diagrams
(b) and (d), as in (c) and (e), comes with an additional factor
of $\delta'$.  Therefore, we must merely make the 
replacement $\delta'\to\delta'/4$ in all but the overall factors of $\delta'$.
This is easily accomplished by letting  $\delta'\to\delta'/4$ in
the computation of the full mass eigenstates (\ie $m^2_{\pi_V^0}$,
$m^2_{\eta_V}$ and $m^2_{\eta'_V}$ in \eq{D_term}), but not in the
overall coefficient of $\cD$.

After making the transition to the $1\!+\!1\!+\!1$ case, taking
the $m_0\to\infty$ limit (with $m^2_{\eta'_I}\sim m_0^2$), 
and using \eq{integrand} through \eq{Integral3}
to perform the momentum integrals, we have
\begin{eqnarray}\label{eq:111_result}
	\delta\! f_{P^+_5} && \to 
	 -\frac{1}{32}\sum_{Q,B} \ell\left(m^2_{Q_B}\right)
	+ \frac{1}{6}\Biggl(R^{[3,3]}_{X_I}(\{\cM^{(1)}_{X_I}\})
	\tilde\ell(m^2_{X_I})
	+R^{[3,3]}_{Y_I}(\{\cM^{(1)}_{Y_I}\})\tilde\ell(m^2_{Y_I})
	 \nonumber \\* &&
	+\sum_{j_I}
	D^{[3,3]}_{j_I,X_I}(\{\cM^{(1)}_{X_I}\})\ell(m^2_{j_I})
	+\sum_{j_I}D^{[3,3]}_{j_I,Y_I}(\{\cM^{(1)}_{Y_I}\})\ell(m^2_{j_I})
	 \nonumber \\* &&
	-2\sum_{j_I}R^{[4,3]}_{j_I}(\{\cM^{(2)}_I\})\ell(m^2_{j_I})\Biggr)
	+\frac{1}{2}a^2 \delta'_V\Biggl[  R^{[4,3]}_{X_V}(\{\cM^{(3)}_{X_V}\})
	\tilde\ell(m^2_{X_V})
	  \nonumber \\* &&
	+   R^{[4,3]}_{Y_V}(\{\cM^{(3)}_{Y_V}\})\tilde\ell(m^2_{Y_V})
	+ \sum_{j_V}D^{[4,3]}_{j_V,X_V}(\{\cM^{(3)}_{X_V}\})\ell(m^2_{j_V})  
	 \nonumber \\* &&
	 +\sum_{j_V}D^{[4,3]}_{j_V,Y_V}(\{\cM^{(3)}_{Y_V}\})\ell(m^2_{j_V})
	 +2\sum_{j_V}R^{[5,3]}_{j_V}(\{\cM^{(4)}_{V}\})\ell(m^2_{j_V})
	 \Biggr]
	+ \Bigl[ V \to A \Bigr]\ ,
\end{eqnarray}
where $Q$ and $B$ have the same meaning as in \eq{df_result}, the
chiral logarithms $\ell(m^2)$ are defined in 
\eqs{chiral_log_infinitev}{chiral_log1} for infinite and finite
spatial volume, respectively, and the $R$s and $D$s are residues
defined in \eqs{residues}{residues2}. The arrow signifies that we are only
keeping the chiral logarithm terms in this expression.
We have defined the sets of masses in the residues:
\begin{eqnarray}\label{eq:denom_mass_sets}
	\{\cM^{(1)}_{Z}\}& \equiv & \{m_{\pi^0},m_{\eta},
		m_{Z}   \}\ , \nonumber \\*
	\{\cM^{(2)}\}& \equiv & \{m_{\pi^0},m_{\eta},
	m_{X},m_{Y}\}\ , \nonumber \\*
	\{\cM^{(3)}_{Z}\}& \equiv & \{m_{\pi^0},m_{\eta},
		m_{\eta'},m_{Z}   \}\ , \nonumber \\*
	\{\cM^{(4)}\}& \equiv & \{m_{\pi^0},m_{\eta},m_{\eta'},
	m_{X},m_{Y}
	   \}\ ,
\end{eqnarray}
where $Z$ can be either $X$ or $Y$, and we show the taste labels explicitly
in \eq{111_result}.  We do not include the numerator masses in the
argument of the residues, as they are the same for each case:
\begin{equation}\label{eq:num_mass_sets}
	\{ \mu \} = \{ m_{U},m_{D},m_{S}   \} \ ,
\end{equation}
with appropriate taste subscripts. The sums over $j_I$, $j_V$,
and $j_A$ run over the set of masses included as the argument of the
residues in each sum.  

The values of $m^2_{\pi^0}$, $m^2_{\eta}$ and $m^2_{\eta'}$
in each taste channel in \eq{111_result} are the eigenvalues of
the full mass matrix
\begin{equation}\label{eq:mass_matrix}
        \left(\matrix{m^2_U +\delta'/4 & \delta'/4 & \delta'/4 \cr
        \delta'/4 & m^2_D +\delta'/4 & \delta'/4 \cr
        \delta'/4 & \delta'/4 & m^2_S +\delta'/4 \cr}\right) \ ,
\end{equation}
where $\delta'$ is given by \eq{dp_def}, the 
masses $m^2_U$, $m^2_D$, $m^2_S$ have an implicit taste label
($V$, $A$, or $I$) depending on the channel, and the
$m^2_0\to\infty$ limit should be taken in the singlet channel.
The explicit expressions for these eigenvalues are not illuminating
in general, but they are given for $m_u=m_d$ (the $2\!+\!1$ case)
in Ref.~\cite{AUBIN}.

In writing down  \eq{111_result}, we
have assumed $m_X^2\ne m_Y^2$. When
$m_X^2= m_Y^2$ some of the residues of sets $\{\cM^{(2)}\}$ and $\{\cM^{(4)}\}$
blow up, so the limit must be taken carefully.  Alternatively,
one can simply return to \eq{df_result}, take the limit 
trivially, and perform the integrations again.

In the quenched case, we drop the first term in \eq{df_result} 
(with the sum over $Q$ and $B$) since it comes from diagrams with
a sea quark loop. In the remaining expression, we make the replacement
$\cD \to \cD^{\, \rm quench}$. 

Writing out the residues explicitly, and using the results from the
Appendix, we obtain
\begin{eqnarray}\label{eq:df_result_qu}
	\delta\! f^{\rm quench}_{P^+_5} && \to 
	\frac{m_0^2}{6} \left[\tilde\ell(m^2_{X_I})
	+\tilde\ell(m^2_{Y_I}) - 2 
	\frac{\ell(m^2_{X_I}) -
	\ell(m^2_{Y_I})}{m^2_{Y_I} - m^2_{X_I}}\right]
	\nonumber \\* &&
	+\frac{\alpha}{6} \left[
	  \ell(m^2_{X_I}) -m^2_{X_I}\tilde\ell(m^2_{X_I})
	  +\ell(m^2_{Y_I}) -m^2_{Y_I}\tilde\ell(m^2_{Y_I})
	  + 2 \frac{m^2_{X_I}\ell(m^2_{X_I}) -
	m^2_{Y_I}\ell(m^2_{Y_I})}{m^2_{Y_I} - m^2_{X_I}}\right]
	\nonumber \\* &&+ \frac{1}{2}a^2\delta'_V
	\Biggl[\tilde\ell(m^2_{X_V})
	+\tilde\ell(m^2_{Y_V}) + 2 
	\frac{\ell(m^2_{X_V}) -
	\ell(m^2_{Y_V})	}{m^2_{Y_V} - m^2_{X_V}}\Biggr]
	+ \Bigl[ V \to A \Bigr] \ .
\end{eqnarray}
Carefully taking the limit $m_y\to m_x$ (or returning to \eq{df_result}
and taking the limit trivially), we see that the singlet terms vanish but the
vector and axial terms do not. This is consistent with the known
result \cite{SHARPE_QCHPT,CBMG_QCHPT} in the symmetry (continuum)
limit that there are no chiral logarithms in
the quenched pion decay constant with degenerate masses.

\section{Final NLO results}
\label{sec:final_results}
For the complete expression for the NLO  pion decay constant, we need
the ``$\cO (p^4)$'' analytic terms in addition to the chiral logarithms
calculated above.  Ours is a joint expansion in $a^2$ and
the quark mass $m$, so we
are looking for the analytic contributions arising from
terms of $\cO(m^2,ma^2,a^4)$ in the chiral Lagrangian. 
Examples of such terms are:
$\Tr(\partial_{\mu}\Sigma\partial_{\mu}\Sigma^{\dagger})\Tr(\cM\Sigma
+ \cM\Sigma^{\dagger})$ [$\cO (m^2)$], 
$a^2\cV\Tr(\partial_{\mu}\Sigma\partial_{\mu}\Sigma^{\dagger})$ 
[$\cO (m a^2)$], and
$(a^2\cV)^2$ [$\cO(a^4)$].
There will also be chiral representatives of those $\cO(a^2)$ operators
in the effective continuum QCD action (``Symanzik action'') that have no
representatives at lowest order --- such operators comprise what
Lee and Sharpe \cite{LEE_SHARPE} call $S_6^{FF(B)}$.
Chiral representatives of $S_6^{FF(B)}$ operators have two 
derivatives \cite{LEE_SHARPE} and therefore are $\cO(ma^2)$. 

The only terms in the chiral Lagrangian that can
contribute to the decay constant are those with 
derivatives, so $\cO(a^4)$ terms are not relevant here. 
Similarly, the ``$m$'' in $\cO (m a^2)$ terms
must come from two derivatives.  Therefore such terms just 
make a NLO contribution to $f_{P^+_5}$ of the form $Fa^2 f$, where
$F$ is a constant formed out
of the coefficients of the relevant Lagrangian terms. $F$ would
of course depend on the taste of the decaying particle, but we
are considering only Goldstone particles here. Finally, the terms
in the Lagrangian that are $\cO (m^2)$ are just the NLO ones familiar 
from continuum \chpt\ \cite{GL_SMALL-M}. 

\subsection{Full and partially quenched NLO results}\label{subsec:pq}

Using the definitions of $L_i$ in Ref.~\cite{GL_SMALL-M},
we thus get from \eqs{111_result}{fpi}, in the 
$1\!+\!1\!+\!1$ partially quenched case,
\begin{eqnarray}\label{eq:final_111_result}
	f^{\rm 1-loop, 1\!+\!1\!+\!1}_{P^+_5} & = &f\Biggl\{ 
	1 + \frac{1}{16\pi^2 f^2} \Biggl[
	  -\frac{1}{32}\sum_{Q,B} \ell\left(m^2_{Q_B}\right)
	  + \frac{1}{6}\Biggl(R^{[3,3]}_{X_I}(\{\cM^{(1)}_{X_I}\})
	  \tilde\ell(m^2_{X_I})      \nonumber \\* &&
	+R^{[3,3]}_{Y_I}(\{\cM^{(1)}_{Y_I}\})\tilde\ell(m^2_{Y_I})
        +\sum_{j_I}
        D^{[3,3]}_{j_I,X_I}(\{\cM^{(1)}_{X_I}\})\ell(m^2_{j_I})
        \nonumber \\* &&
       +\sum_{j_I}D^{[3,3]}_{j_I,Y_I}(\{\cM^{(1)}_{Y_I}\})\ell(m^2_{j_I})
          -2\sum_{j_I}R^{[4,3]}_{j_I}(\{\cM^{(2)}_I\})\ell(m^2_{j_I})\Biggr)
         \nonumber \\* &&
        +\frac{1}{2}a^2 \delta'_V\Biggl(  R^{[4,3]}_{X_V}(\{\cM^{(3)}_{X_V}\})
        \tilde\ell(m^2_{X_V})
        +   R^{[4,3]}_{Y_V}(\{\cM^{(3)}_{Y_V}\})\tilde\ell(m^2_{Y_V})
         \nonumber \\* &&
       + \sum_{j_V}D^{[4,3]}_{j_V,X_V}(\{\cM^{(3)}_{X_V}\})\ell(m^2_{j_V})
          +\sum_{j_V}D^{[4,3]}_{j_V,Y_V}(\{\cM^{(3)}_{Y_V}\})\ell(m^2_{j_V})
          \nonumber \\* &&
	  +2\sum_{j_V}R^{[5,3]}_{j_V}(\{\cM^{(4)}_{V}\})\ell(m^2_{j_V})
	 \Biggr) + \Bigl( V \to A \Bigr) \Biggr]\nonumber \\* &&
	+ \frac{16\mu}{f^2}\left( m_u + m_d + m_s\right)L_4
	+ \frac{8\mu}{f^2}\left( m_x + m_y \right)L_5
	+a^2 F \Biggr\} \ .
\end{eqnarray}
Definitions here are the same as in \eq{111_result}.
 We have checked that this result
reduces to that of Sharpe and Shoresh \cite{UNPHYSICAL}
in the continuum (symmetry) limit. Using \eq{identities}, it is not
hard to show that changes here in the chiral scale 
$\Lambda$ can be absorbed
into the parameters $L_4$, $L_5$ and $F$, as expected.

In the $2\!+\!1$ case ($m_u=m_d\equiv
m_{\ell}$) with no other degeneracies, there is some simplification
because $m^2_{\pi^0}=m^2_U=m^2_D$ in each taste channel. We obtain:
\begin{eqnarray}\label{eq:final_21_result}
	f^{\rm 1-loop, 2\!+\!1}_{P^+_5} && = f\Biggl\{ 1 
	+ \frac{1}{16\pi^2 f^2}
	 \Biggl[-\frac{1}{32}\sum_{Q,B} \ell\left(m^2_{Q_B}\right)
	+ \frac{1}{6}\Biggl(R^{[2,2]}_{X_I}(\{\cM^{(5)}_{X_I}\})
	\tilde\ell(m^2_{X_I})
	 \nonumber \\* &&
	+R^{[2,2]}_{Y_I}(\{\cM^{(5)}_{Y_I}\})\tilde\ell(m^2_{Y_I})
	+\sum_{j_I}
	D^{[2,2]}_{j_I,X_I}(\{\cM^{(5)}_{X_I}\})\ell(m^2_{j_I})
	+\sum_{j_I}D^{[2,2]}_{j_I,Y_I}(\{\cM^{(5)}_{Y_I}\})\ell(m^2_{j_I})
	 \nonumber \\* &&
	-2\sum_{j_I}R^{[3,2]}_{j_I}(\{\cM^{(6)}_I\})\ell(m^2_{j_I})\Biggr)
	+\frac{1}{2}a^2 \delta'_V\Biggl(  R^{[3,2]}_{X_V}(\{\cM^{(7)}_{X_V}\})
	\tilde\ell(m^2_{X_V})
	  \nonumber \\* &&
	+   R^{[3,2]}_{Y_V}(\{\cM^{(7)}_{Y_V}\})\tilde\ell(m^2_{Y_V})
	+ \sum_{j_V}  D^{[3,2]}_{j_V,X_V}(\{\cM^{(7)}_{X_V}\})\ell(m^2_{j_V})
	  \nonumber \\* &&
	+\sum_{j_V}D^{[3,2]}_{j_V,Y_V}(\{\cM^{(7)}_{Y_V}\})\ell(m^2_{j_V})
	+2\sum_{j_V}R^{[4,2]}_{j_V}
	(\{\cM^{(8)}_{V}\})\ell(m^2_{j_V})
	 \Biggr)	+ \Bigl( V \to A \Bigr) \Biggr]
	  \nonumber \\* &&
	+ \frac{16\mu}{f^2}\left( 2m_{\ell} + m_s\right)L_4
 	+ \frac{8\mu}{f^2}\left( m_x + m_y \right)L_5
	+a^2 F \Biggr\} \ ,
\end{eqnarray}
with definitions the same as in \eq{111_result}, except that now the
denominator masses in the residues are:
\begin{eqnarray}\label{eq:denom_mass_sets2}
	\{\cM^{(5)}_{Z}\}& \equiv & \{m_{\eta},
		m_{Z}   \}\ , \nonumber \\*
	\{\cM^{(6)}\}& \equiv & \{m_{\eta},
	m_{X},m_{Y}\}\ , \nonumber \\*
	\{\cM^{(7)}_{Z}\}& \equiv & \{m_{\eta},
		m_{\eta'},m_{Z}   \}\ , \nonumber \\*
	\{\cM^{(8)}\}& \equiv & \{m_{\eta},m_{\eta'},
	m_{X},m_{Y}
	   \}\ ,
\end{eqnarray}
where $Z$ can again be either $X$ or $Y$, and a taste label is
implicit.  The numerator masses in the
residues of \eq{final_21_result} are not shown explicitly.  They
are always \begin{equation}\label{eq:num_mass_sets2}
        \{ \mu \} = \{ m_{U},m_{S}   \} \ ,
\end{equation}
with the taste label again implicit. 

Various cases of interest can be obtained either by carefully
taking limits in 
\eqsor{final_111_result}{final_21_result}
, or 
by taking the limits in \eq{df_result}  and redoing 
the momentum integration.
We first consider the ``full QCD'' case of ``real'' pions
and kaons. By setting $m_x=m_u$
and $m_y=m_d$, but keeping $m_u\ne m_d$,
we get after a bit of algebra:
\begin{eqnarray}\label{eq:final_111_pion_result}
	f^{\rm 1-loop, 1\!+\!1\!+\!1}_{\pi^+_5} & \!\!= & f\Biggl\{ 1 
	+ \frac{1}{16\pi^2 f^2}
	 \Biggl[ -\frac{1}{32}\sum_{Q,B}  \ell\left(m^2_{Q_B}\right)
\nonumber \\* &&
        + \frac{(m^2_{U_I} - m^2_{D_I})^2}{6}
         \sum_{j_I}R^{[4,1]}_{j_I}\left(\{\cM^{(2)}_I\};\{m^2_{S_I}\}\right)
        \ell(m^2_{j_I}) \nonumber \\* &&
        \!\!+ \frac{1}{2}a^2\delta'_V\biggl( \sum_{j_V}
            (m^2_{U_V} + m^2_{D_V} - 2m^2_{j_V})^2\;
         R^{[5,1]}_{j_V}\left(\{\cM^{(4)}_V\};\{m^2_{S_V}\}\right)
           \ell(m^2_{j_V})  
        \biggr)  \nonumber \\* &&
        \!\!+ \Bigl(V\to A  \Bigr)\Biggr]
	+ \frac{16\mu}{f^2}\left( m_u + m_d + m_s\right)L_4
	+ \frac{8\mu}{f^2}\left( m_u + m_d \right)L_5
	+a^2 F \Biggr\}.
\end{eqnarray}
where the sets $\{\cM^{(2)}\}$ and $\{\cM^{(4)}\}$ are given in
\eq{denom_mass_sets} (with $X\to U$ and $Y\to D$), 
$j_I$ and $j_V$ run over all masses in $\{\cM_I^2\}$ and $\{\cM_V^4\}$, 
respectively, and
$Q\in \{ U, D, \pi^+, \pi^-, K^+, K^0 \}$. Note that
there are no double pole terms here, due to cancellations in the
disconnected flavor-neutral propagator. The charged kaon result can be
obtained from \eq{final_111_pion_result} by 
making the replacements $d \leftrightarrow s$ and $D
\leftrightarrow S$ wherever they appear explicitly (as well as
in the definitions of the mass sets $\{\cM_I^2\}$ and $\{\cM_V^4\}$ and in the
sum over $Q$, where we now have
$Q\in \{ U, S, K^+, K^-, \pi^+, \bar K^0 \}$).

The full theory pion result simplifies even more in the $2\!+\!1$
case, when the up
and down quark masses are equal. After writing the residues
explicitly, we obtain:
\begin{eqnarray}\label{eq:final_21_pion_result}
	f^{\rm 1-loop, 2\!+\!1}_{\pi^+_5} && = f\Biggl\{ 1 + 
	\frac{1}{16\pi^2 f^2}
	 \Biggl[-\frac{1}{16}\sum_{B}\left(
	  2\ell(m^2_{\pi^0_B})+ \ell(m^2_{K^+_B})\right)
	 \nonumber \\* &&
	\!\!\!+ 2a^2\delta'_V\Biggl(  \frac{m^2_{S_V}-m^2_{\eta_V}}
	{(m^2_{\pi^0_V}-m^2_{\eta_V})(m^2_{\eta'_V}-m^2_{\eta_V})}
	\ell(m^2_{\eta_V})
	+ \frac{m^2_{S_V}-m^2_{\eta'_V}}
	{(m^2_{\pi^0_V}-m^2_{\eta'_V})(m^2_{\eta_V}-m^2_{\eta'_V})}
	\ell(m^2_{\eta'_V})\nonumber \\* && 
	+  \frac{m^2_{S_V}-m^2_{\pi^0_V}}
	{(m^2_{\eta_V}-m^2_{\pi^0_V})(m^2_{\eta'_V}-m^2_{\pi^0_V})}
	\ell(m^2_{\pi^0_V}) \Biggr)
	+ \Bigl( V\to A  \Bigr)\Biggr] \nonumber\\*&&
	+ \frac{16\mu}{f^2}\left( 2m_{\ell} + m_s\right)L_4
	+ \frac{16\mu}{f^2}m_{\ell} L_5
	+a^2 F \Biggr\}  \ .
\end{eqnarray}

Similarly, the $2\!+\!1$ result for the full kaon is:
\begin{eqnarray}\label{eq:final_21_kaon_result}
	f^{\rm 1-loop, 2\!+\!1}_{K^+_5} && = f\Biggl\{ 1 + 
	\frac{1}{16\pi^2 f^2}
	 \Biggl[-\frac{1}{32}\sum_{B}\left(
	  2\ell(m^2_{\pi^0_B})+ 3\ell(m^2_{K^+_B}) +  \ell(m^2_{S_B})\right)
	 \nonumber \\* &&
	+\frac{1}{4}\ell(m^2_{\pi^0_I}) 
	-\frac{3}{4}\ell(m^2_{\eta_I}) +\frac{1}{2}\ell(m^2_{S_I})
	\nonumber \\* &&
	\!\!\!+ \frac{1}{2}a^2\delta'_V\Biggl(  
         \frac{(m^2_{S_V}+m^2_{\pi^0_V}-2m^2_{\eta_V})^2}
	{(m^2_{\pi^0_V}-m^2_{\eta_V})(m^2_{S_V}-m^2_{\eta_V})(m^2_{\eta'_V}-m^2_{\eta_V})}
	\ell(m^2_{\eta_V}) \nonumber\\*&&
         +\frac{(m^2_{S_V}+m^2_{\pi^0_V}-2m^2_{\eta'_V})^2}
	{(m^2_{\pi^0_V}-m^2_{\eta'_V})(m^2_{S_V}-m^2_{\eta'_V})(m^2_{\eta_V}-m^2_{\eta'_V})}
	\ell(m^2_{\eta'_V}) \nonumber\\*&&
	+  \frac{m^2_{S_V}-m^2_{\pi^0_V}}
	{(m^2_{\eta_V}-m^2_{\pi^0_V})(m^2_{\eta'_V}-m^2_{\pi^0_V})}
	\ell(m^2_{\pi^0_V}) 
	+  \frac{m^2_{\pi^0_V}-m^2_{S_V}}
	{(m^2_{\eta_V}-m^2_{S_V})(m^2_{\eta'_V}-m^2_{S_V})}
	\ell(m^2_{S_V}) \Biggr)  \nonumber \\* &&
	+ \Bigl( V\to A  \Bigr)\Biggr] 
	+ \frac{16\mu}{f^2}\left( 2m_{\ell} + m_s\right)L_4
	+ \frac{8\mu}{f^2}(m_{\ell} +m_s)L_5
	+a^2 F \Biggr\}  \ .
\end{eqnarray}
Here we have used the fact that $m^2_{\eta_I}=\frac{2}{3}m^2_{S_I} +
\frac{1}{3} m^2_{\pi^0_I}$ in the $2\!+\!1$ case to simplify the result.
It is easy to check that \eqs{final_21_pion_result}{final_21_kaon_result}
reduce to the standard answers \cite{GL_SMALL-M} in the 
$a^2\to0$ limit, where all tastes are degenerate.

\subsection{Quenched NLO Results}\label{subsec:qu}

In the fully quenched case, we only need to consider the two cases
$m_x\ne m_y$ and $m_x = m_y$.

For the quenched ``kaon'' case ($m_x\ne m_y$) we obtain:
\begin{eqnarray}\label{eq:final_result_qu}
	f^{\rm 1-loop, quench}_{K^+_5} & = &f\Biggl\{ 1 + \frac{1}{16\pi^2 f^2}
	 \Biggl[\frac{m_0^2}{6} \left(\tilde\ell(m^2_{X_I})
	+\tilde\ell(m^2_{Y_I}) - 2 
	\frac{\ell(m^2_{X_I}) -
	\ell(m^2_{Y_I})}{m^2_{Y_I} - m^2_{X_I}}\right)
	  \nonumber \\* &&+\frac{\alpha}{6} \Biggl(
	  \ell(m^2_{X_I}) -m^2_{X_I}\tilde\ell(m^2_{X_I})
	  +\ell(m^2_{Y_I}) -m^2_{Y_I}\tilde\ell(m^2_{Y_I})
	  \nonumber \\*&&+ 2 \frac{m^2_{X_I}\ell(m^2_{X_I}) -
	m^2_{Y_I}\ell(m^2_{Y_I})}{m^2_{Y_I} - m^2_{X_I}}\Biggr)
	+ \frac{1}{2}a^2\delta'_V
	\Biggl(\tilde\ell(m^2_{X_V})
	+\tilde\ell(m^2_{Y_V}) \nonumber \\*&&+ 2 
	\frac{\ell(m^2_{X_V}) -
	\ell(m^2_{Y_V})	}{m^2_{Y_V} - m^2_{X_V}}\Biggr)
	+ \Bigl( V \to A \Bigr) \Biggr] 
	+ \frac{8\mu}{f^2}\left( m_x + m_y \right)L_5'
	+a^2 F' \Biggr\}.
\end{eqnarray}
The analytic terms in the quenched case are marked with primes to
indicate that they may have different values than in the full
theory. Also, note that there is no analytic term involving the sea
quarks in the quenched case, as they play no role here.
In the continuum limit, \eq{final_result_qu} reproduces the
known quenched result \cite{CBMG_QCHPT}.

Taking the degenerate limit ($m_y = m_x$)
in the quenched case, we obtain for the quenched ``pion'':
\begin{equation}\label{eq:final_pion_result_qu}
	f^{\rm 1\!-\!loop, quench}_{\pi^+_5} \! =\! 
	f\Biggl\{ 1 \!+\! \frac{1}{16\pi^2 f^2}
	 \left(2a^2\delta'_V\tilde\ell(m^2_{X_V})
	 + 2a^2\delta'_A\tilde\ell(m^2_{X_A}) \right)
	+ \frac{16\mu}{f^2} m_{x}L_5'
	+a^2 F' \Biggr\}.
\end{equation}
This is consistent with the fact that in the isospin limit, the
continuum quenched pion decay constant does not contain chiral logarithms.

\section{Remarks and Conclusions}\label{sec:conc}
The most general result we have is for the $n = 3$ partially quenched
case ($1\!+\!1\!+\!1$) with all valence and sea quark masses different,
\eq{final_111_result}. Other interesting cases can be obtained from
\eq{final_111_result} by taking appropriate mass limits. The results
most relevant to current MILC simulations are those with
$m_u=m_d\equiv m_l$ (the $2\!+\!1$ case); these and other important limits
are presented explicitly in 
Sec.~\ref{subsec:pq}.
The results
in the quenched case are given separately in 
Sec.~\ref{subsec:qu}, in \eqs{final_result_qu}{final_pion_result_qu}.

The explicit results in Sec.~\ref{sec:final_results} often
appear dauntingly complex.  However, the intricacies arise primarily from
the momentum integration, which produces chiral logarithms with 
complicated residues from each of the many poles in
the disconnected flavor neutral propagator, \eq{D_term}.
The result before integration, \eq{df_result}, is actually quite simple,
and the reader may prefer to start with that expression and perform
the integration himself in specific cases of interest.

In the partially quenched case, double poles arise here even when
the valence masses are non-degenerate, just as they do in the
continuum \cite{CBMG_PQCHPT,UNPHYSICAL}.  It is interesting that these
double poles appear in the explicitly $\cO(a^2)$ terms (taste-vector
or axial channels, proportional
to $\delta'_V$ or $\delta'_A$) as well as in the continuum-like
taste-singlet channel.

Using the \schpt\ results presented here and in \cite{AUBIN}, it
seems possible to fit existing lattice data and extract physical physical
parameters (\eg $f_\pi$, $f_K$, $m_s$, ($m_u$+$m_d$)/2, $L_i$)
with rather small discretization errors \cite{MILC_FITS}.
The next steps would be to extend the current approach
to describe heavy-light mesons \cite{AUBIN_BERNARD} and baryons.

\bigskip
\bigskip
\centerline{\bf ACKNOWLEDGMENTS}
\bigskip
We thank our colleagues in the MILC collaboration for 
helpful discussions.  This work was partially supported by the
U.S. Department of Energy under grant number DE-FG02-91ER40628.

\appendix

\section*{Appendix}\label{ap:integral}

In this Appendix, we go through the technical details of calculating
the integrals found in \eq{df_result}. For the terms only containing single
poles, this was done in Ref.~\cite{AUBIN}, so here we focus on the terms
which contain double poles.

Consider first an integrand of the form
\begin{equation}\label{eq:integrand}
\cI^{[n,k]}\left(\left\{m\right\}\!;\!\left\{\mu\right\}\right)
\equiv \frac{\prod_{a=1}^k (q^2 + \mu^2_a)}
{\prod_{j=1}^n (q^2 + m^2_j)} \ ,
\end{equation}
where $\{m\}$ and $\{\mu\}$ are the sets of masses
$\{m_1,m_2,\dots,m_n\}$ and $\{\mu_1,\mu_2,\dots,\mu_k\}$, respectively,
As long as $n>k$ and there are no mass degeneracies in the
denominator, 
$\cI^{[n,k]}$ can be written as the sum of simple poles
times their residues:
\begin{equation}\label{eq:lagrange}
        \cI^{[n,k]}\left(\left\{m\right\}\!;\!\left\{\mu\right\}\right) =
        \sum_{j=1}^n \frac{R_j^{[n,k]}\left(\left\{m\right\}\!;\!\left\{\mu\right\}\right)}{q^2 + m^2_j}\ ,
\end{equation}
where
\begin{equation}\label{eq:residues}
	R_j^{[n,k]}\left(\left\{m\right\}\!;\!\left\{\mu\right\}\right)
	 \equiv  \frac{\prod_{a=1}^k (\mu^2_a- m^2_j)}
	{\prod_{i\not=j} (m^2_i - m^2_j)}\ .
\end{equation}

If there is a double pole, the residues are modified. 
We need consider only the case of one double pole; let it occur
at $q^2 =-m_\ell^2$.
We then have
\begin{eqnarray}\label{eq:integrand2}
	\cI^{[n,k]}_{\rm dp}\left(m_{\ell};\left\{m\right\}\!
	;\!\left\{\mu\right\}\right) 
	&\equiv& \frac{\prod_{a=1}^k (q^2 + \mu^2_a)}
	{(q^2 + m^2_{\ell})\prod_{j=1}^{n} (q^2 + m^2_j)}
	\nonumber\\*
	& = & -\frac{d}{d m^2_{\ell}}\left(
	 \frac{\prod_{a=1}^k (q^2 + \mu^2_a)}
	{\prod_{j=1}^{n} (q^2 + m^2_j)}
	\right) \ .
\end{eqnarray}
Here the product over $j$ includes $\ell$, \ie $1\le\ell\le n$.
We now expand the quantity inside of the derivative as 
a sum of single poles and
take the derivative of the resulting
expression. The result is 
\begin{equation}\label{eq:integrand2_expanded}
	\cI^{[n,k]}_{\rm dp}\left(m_{\ell};\left\{m\right\}\!
	;\!\left\{\mu\right\}\right) 
	= \frac{R_{\ell}^{[n,k]}\left(\left\{m\right\}\!;
	\!\left\{\mu\right\}\right)}{(q^2 + m^2_{\ell})^2}
	+ \sum_{j=1}^{n}\frac{D_{j, \ell}^{[n,k]}
	  \left(\left\{m\right\}\!;
	\!\left\{\mu\right\}\right)}{(q^2 + m^2_{j})} \ ,
\end{equation}
with
\begin{equation}\label{eq:residues2}
	D_{j, \ell}^{[n,k]}\left(\left\{m\right\}\!;
	\!\left\{\mu\right\}\right) \equiv -\frac{d}{d m^2_{\ell}}
	R_{j}^{[n,k]}\left(\left\{m\right\}\!;
	\!\left\{\mu\right\}\right)\ .
\end{equation}
Note that $D_{j, \ell}^{[n,k]}$ takes on a simple form
for $j\ne \ell$: 
\begin{equation}\label{eq:D-j-not-equal-l}
 D_{j, \ell}^{[n,k]} \left(\left\{m\right\}\!;
        \!\left\{\mu\right\}\right) = R_j^{[n+1,k]}\left(\left\{m\right\}'\!;
        \!\left\{\mu\right\}\right)\qquad (j\ne \ell)\ ,
\end{equation}
where $\{m\}'$ is just the set $\{m\}$ with $m_\ell$ repeated:
$\{m\}' = \{m_1,\dots,m_\ell,m_\ell,\dots, m_n\}$.
For $j=\ell$, $D_{j, \ell}^{[n,k]}$ becomes quite
complicated, with $n+k$ terms due to the differentiation. We emphasize
that these formulae are needed solely for performing the momentum integrals
explicitly in the partially quenched case.
In full QCD, there are no double poles.

We now collect some identities satisfied by the residues:
\begin{eqnarray}\label{eq:identities}
\sum_{j=1}^n  R_{j}^{[n,k]} &=& \cases {\phantom{-}1\ , &$n=k+1$ ; \cr
					 \phantom{-}0\ , &$n\ge k+2$ . \cr }
\nonumber \\*
\sum_{j=1}^n  R_{j}^{[n,k]}m^2_j  &= & \cases {\sum_{j=1}^n m^2_j - 
				\sum_{a=1}^k \mu^2_a  \ , &$n=k+1$ ; \cr
					 -1\ , &$n=k+2$ ; \cr 
					 \phantom{-}0\ , &$n\ge k+3$ . \cr }
\nonumber \\*
\sum_{j=1}^n  D_{j,\ell}^{[n,k]} &=& \cases { \phantom{-}1\ , &$n=k$ ; \cr
					 \phantom{-}0\ , &$n\ge k+1$ . \cr }
\nonumber \\*
\sum_{j=1}^n  \left(D_{j,\ell}^{[n,k]}m^2_j\right)
- R_{\ell}^{[n,k]}  &= & \cases {m^2_\ell + \sum_{j=1}^n m^2_j - 
				\sum_{a=1}^k \mu^2_a  \ , &$n=k$ ; \cr
					 -1\ , &$n=k+1$ ; \cr 
					  \phantom{-}0\ , &$n\ge k+2$ . \cr }
\end{eqnarray}
These identities are easily obtained by expanding both sides of 
\eqsor{lagrange}{integrand2_expanded} for large $q^2$.

When performing the explicit evaluation,
the following integrals are needed:
\begin{eqnarray}\label{eq:Integral1}
	\cI_1& \equiv& \int \frac{d^4 q}{(2\pi)^4}
	 \frac{1}{q^2 + m^2} \rightarrow \frac{1}{16\pi^2} 
	\ell( m^2)\ , \\*
	 \label{eq:Integral3}
	\cI_2& \equiv& \int \frac{d^4 q}{(2\pi)^4}
	 \frac{q^2}{q^2 + m^2} = \int\frac{d^4 q}{(2\pi)^4} - 
	 m^2\cI_1 \rightarrow \frac{-m^2}{16\pi^2} 
	 \ell( m^2)\ , \\*
	\cI_3 &\equiv& \int \frac{d^4 q}{(2\pi)^4}
	 \frac{1}{(q^2 + m^2)^2} = -\frac{\partial}
	{\partial m^2}\cI_1\rightarrow
	\frac{1}{16\pi^2}  \tilde \ell(m^2) \ , \\*
	 \label{eq:Integral4}
	\cI_4 &\equiv& \int \frac{d^4 q}{(2\pi)^4}
	 \frac{q^2}{(q^2 + m^2)^2} = 
	 \cI_1 - m^2\cI_3 \rightarrow
	 \frac{1}{16\pi^2} \left( \ell(m^2)-m^2\tilde \ell(m^2) \right)\ ,
\end{eqnarray}
where we have defined the chiral logarithm functions
\begin{eqnarray}\label{eq:chiral_log_infinitev}
	\ell(m^2) &\equiv & m^2 \ln \frac{m^2}{\Lambda^2}
	\qquad{\rm [infinite\ volume]} \ , \\*
	\label{eq:chiral_log2_infinitev}
	\tilde \ell(m^2)& \equiv&  -\left(\ln 
	\frac{m^2}{\Lambda^2} + 1\right)\qquad
	{\rm [infinite\ volume]} \ ,
\end{eqnarray}
with $\Lambda$ the chiral scale. We use the arrow in
 \eqsthru{Integral1}{Integral4}
and elsewhere to indicate that we are only keeping the chiral logarithm
terms. If the system is in a finite (but large) spatial volume $L^3$,
the following modifications are required \cite{CHIRAL_FSB}:
\begin{eqnarray}\label{eq:chiral_log1}
	\ell( m^2)& \equiv & m^2 \left(\ln \frac{m^2}{\Lambda^2}
	 + \delta_1(mL)\right) \qquad{\rm [finite\ spatial\ volume]} \ ,
	\\*
	\label{eq:chiral_log2}
	 \tilde \ell(m^2)& \equiv & -\left(\ln 
	\frac{m^2}{\Lambda^2} + 1\right) + \delta_3(mL) 
	\qquad{\rm [finite\ spatial\ volume]} \ ,
\end{eqnarray}
where
\begin{eqnarray}\label{eq:delta1}
	\delta_1(mL) & = & \frac{4}{mL}
		\sum_{\vec r\ne 0}
		\frac{K_1(|\vec r|mL)}{|\vec r|} \ , \\*
	\label{eq:delta3}
	\delta_3(mL) & =& 2 \sum_{\vec r\ne 0}
		K_0(|\vec r|mL)\ ,
\end{eqnarray}
with $K_0$ and $K_1$ the Bessel functions of imaginary argument.

\vfill\eject


\end{document}